\documentclass[aps,prl,preprint,superscriptaddress]{revtex4-1}

\usepackage{graphicx}
\usepackage{dcolumn}
\usepackage[normalem]{ulem}

\def\add{}

\begin{document}


\title{Gamma-Ray Localization of Terrestrial Gamma-Ray Flashes}

\author{M.~Marisaldi}
\affiliation{INAF-IASF Bologna, Via Gobetti 101, I-40129 Bologna, Italy}
\author{A.~Argan}
\affiliation{INAF, Viale del Parco Mellini 84, Roma, Italy}
\author{A.~Trois}
\affiliation{INAF-IASF Roma, via del Fosso del Cavaliere 100, I-00133 Roma, Italy}
\author{A.~Giuliani}
\affiliation{INAF-IASF Milano, via E. Bassini 15, I-20133 Milano, Italy}
\author{M.~Tavani}
\affiliation{INAF-IASF Roma, via del Fosso del Cavaliere 100, I-00133 Roma, Italy}
\affiliation{Dipartimento di Fisica, Universit\`a Tor Vergata, via della Ricerca Scientifica 1, I-00133 Roma, Italy}
\author{C.~Labanti}
\affiliation{INAF-IASF Bologna, Via Gobetti 101, I-40129 Bologna, Italy}
\author{F.~Fuschino}
\affiliation{INAF-IASF Bologna, Via Gobetti 101, I-40129 Bologna, Italy}
\author{A.~Bulgarelli}
\affiliation{INAF-IASF Bologna, Via Gobetti 101, I-40129 Bologna, Italy}
\author{F.~Longo}
\affiliation{Dipartimento di Fisica Universit\`a di Trieste, via A. Valerio 2, I-34127 Trieste, Italy}

\author{G.~Barbiellini}
\affiliation{Dipartimento di Fisica Universit\`a di Trieste, via A. Valerio 2, I-34127 Trieste, Italy}
\affiliation{INFN Trieste, via A. Valerio 2, I-34127 Trieste, Italy}
\author{E.~Del~Monte}
\affiliation{INAF-IASF Roma, via del Fosso del Cavaliere 100, I-00133 Roma, Italy}
\author{E.~Moretti}
\affiliation{Dipartimento di Fisica Universit\`a di Trieste, via A. Valerio 2, I-34127 Trieste, Italy}
\author{M.~Trifoglio}
\affiliation{INAF-IASF Bologna, Via Gobetti 101, I-40129 Bologna, Italy}

\author{E.~Costa}
\affiliation{INAF-IASF Roma, via del Fosso del Cavaliere 100, I-00133 Roma, Italy}
\author{P.~Caraveo}
\affiliation{INAF-IASF Milano, via E. Bassini 15, I-20133 Milano, Italy}
\author{P.W.~Cattaneo}
\affiliation{INFN Pavia, via Bassi 6, I-27100 Pavia, Italy}
\author{A.~Chen}
\affiliation{INAF-IASF Milano, via E. Bassini 15, I-20133 Milano, Italy}
\author{F.~D'Ammando}
\affiliation{INAF-IASF Palermo, Via Ugo La Malfa 153, 90146 Palermo, Italy}
\author{G.~De~Paris}
\affiliation{INAF, Viale del Parco Mellini 84, Roma, Italy}
\author{G.~Di~Cocco}
\affiliation{INAF-IASF Bologna, Via Gobetti 101, I-40129 Bologna, Italy}
\author{G.~Di~Persio}
\affiliation{INAF-IASF Roma, via del Fosso del Cavaliere 100, I-00133 Roma, Italy}
\author{I.~Donnarumma}
\affiliation{INAF-IASF Roma, via del Fosso del Cavaliere 100, I-00133 Roma, Italy}
\author{Y.~Evangelista}
\affiliation{INAF-IASF Roma, via del Fosso del Cavaliere 100, I-00133 Roma, Italy}
\author{M.~Feroci}
\affiliation{INAF-IASF Roma, via del Fosso del Cavaliere 100, I-00133 Roma, Italy}
\author{A.~Ferrari}
\affiliation{Dipartimento di Fisica, Universit\`a Torino, Torino, Italy}
\affiliation{CIFS Torino, Viale Settimio Severo 63, I-10133 Torino, Italy}
\author{M.~Fiorini}
\affiliation{INAF-IASF Milano, via E. Bassini 15, I-20133 Milano, Italy}
\author{T.~Froysland}
\affiliation{Dipartimento di Fisica, Universit\`a Tor Vergata, via della Ricerca Scientifica 1, I-00133 Roma, Italy}
\author{M.~Galli}
\affiliation{ENEA, via Martiri di Monte Sole 4, I-40129 Bologna, Italy}
\author{F.~Gianotti}
\affiliation{INAF-IASF Bologna, Via Gobetti 101, I-40129 Bologna, Italy}
\author{I.~Lapshov}
\affiliation{IKI, Moscow, Russia}
\author{F.~Lazzarotto}
\affiliation{INAF-IASF Roma, via del Fosso del Cavaliere 100, I-00133 Roma, Italy}
\author{P.~Lipari}
\affiliation{Dipartimento di Fisica, Universit\`a La Sapienza, p.le Aldo Moro 2, I-00185 Roma, Italy}
\affiliation{INFN Roma ``La Sapienza'', p.le Aldo Moro 2, I-00185 Roma, Italy}
\author{S.~Mereghetti}
\affiliation{INAF-IASF Milano, via E. Bassini 15, I-20133 Milano, Italy}
\author{A.~Morselli}
\affiliation{INFN Roma ``Tor Vergata'', via della Ricerca Scientifica 1, I-00133 Roma, Italy}
\author{L.~Pacciani}
\affiliation{INAF-IASF Roma, via del Fosso del Cavaliere 100, I-00133 Roma, Italy}
\author{A.~Pellizzoni}
\affiliation{INAF-Osservatorio Astronomico di Cagliari, loc. Poggio dei Pini, strada 54, I-09012, Capoterra (CA), Italy}
\author{F.~Perotti}
\affiliation{INAF-IASF Milano, via E. Bassini 15, I-20133 Milano, Italy}
\author{P.~Picozza}
\affiliation{Dipartimento di Fisica, Universit\`a Tor Vergata, via della Ricerca Scientifica 1, I-00133 Roma, Italy}
\affiliation{INFN Roma ``Tor Vergata'', via della Ricerca Scientifica 1, I-00133 Roma, Italy}
\author{G.~Piano}
\affiliation{INAF-IASF Roma, via del Fosso del Cavaliere 100, I-00133 Roma, Italy}
\affiliation{Dipartimento di Fisica, Universit\`a Tor Vergata, via della Ricerca Scientifica 1, I-00133 Roma, Italy}
\affiliation{INFN Roma ``Tor Vergata'', via della Ricerca Scientifica 1, I-00133 Roma, Italy}
\author{M.~Pilia}
\affiliation{INAF-Osservatorio Astronomico di Cagliari, loc. Poggio dei Pini, strada 54, I-09012, Capoterra (CA), Italy}
\affiliation{Dipartimento di Fisica, Universit\`a dell'Insubria, Via Valleggio 11, I-22100 Como, Italy}
\author{M.~Prest}
\affiliation{Dipartimento di Fisica, Universit\`a dell'Insubria, Via Valleggio 11, I-22100 Como, Italy}
\author{G.~Pucella}
\affiliation{ENEA Frascati, via Enrico Fermi 45, I-00044 Frascati(Roma), Italy}
\author{M.~Rapisarda}
\affiliation{ENEA Frascati, via Enrico Fermi 45, I-00044 Frascati(Roma), Italy}
\author{A.~Rappoldi}
\affiliation{INFN Pavia, via Bassi 6, I-27100 Pavia, Italy}
\author{A.~Rubini}
\affiliation{INAF-IASF Roma, via del Fosso del Cavaliere 100, I-00133 Roma, Italy}
\author{S.~Sabatini}
\affiliation{INAF-IASF Roma, via del Fosso del Cavaliere 100, I-00133 Roma, Italy}
\author{P.~Soffitta}
\affiliation{INAF-IASF Roma, via del Fosso del Cavaliere 100, I-00133 Roma, Italy}
\author{E.~Striani}
\affiliation{INAF-IASF Roma, via del Fosso del Cavaliere 100, I-00133 Roma, Italy}
\author{E.~Vallazza}
\affiliation{INFN Trieste, via A. Valerio 2, I-34127 Trieste, Italy}
\author{S.~Vercellone}
\affiliation{INAF-IASF Palermo, Via Ugo La Malfa 153, 90146 Palermo, Italy}
\author{V.~Vittorini}
\affiliation{INAF-IASF Roma, via del Fosso del Cavaliere 100, I-00133 Roma, Italy}
\author{A.~Zambra}
\affiliation{INAF, Osservatorio Astronomico di Brera, via Brera 28, 20121 Milano}
\author{D.~Zanello}
\affiliation{INFN Roma ``La Sapienza'', p.le Aldo Moro 2, I-00185 Roma, Italy}

\author{L.A.~Antonelli}
\affiliation{ASI Science Data Center, Via E. Fermi 45, I-00044 Frascati (Roma), Italy}
\author{S.~Colafrancesco}
\affiliation{ASI Science Data Center, Via E. Fermi 45, I-00044 Frascati (Roma), Italy}
\author{S.~Cutini}
\affiliation{ASI Science Data Center, Via E. Fermi 45, I-00044 Frascati (Roma), Italy}
\author{P.~Giommi}
\affiliation{ASI Science Data Center, Via E. Fermi 45, I-00044 Frascati (Roma), Italy}
\author{F.~Lucarelli}
\affiliation{ASI Science Data Center, Via E. Fermi 45, I-00044 Frascati (Roma), Italy}
\author{C.~Pittori}
\affiliation{ASI Science Data Center, Via E. Fermi 45, I-00044 Frascati (Roma), Italy}
\author{P.~Santolamazza}
\affiliation{ASI Science Data Center, Via E. Fermi 45, I-00044 Frascati (Roma), Italy}
\author{F.~Verrecchia}
\affiliation{ASI Science Data Center, Via E. Fermi 45, I-00044 Frascati (Roma), Italy}
\author{L.~Salotti}
\affiliation{Agenzia Spaziale Italiana, viale Liegi 26, I-00198 Roma, Italy}

\date{\today}


\begin{abstract}


Terrestrial Gamma-Ray Flashes (TGFs) are very  short bursts of  high energy photons and electrons originating in 
Earth's atmosphere. We present here a localization study of TGFs carried out at
gamma-ray energies above 20 MeV based on an innovative event selection method.
We use the AGILE satellite Silicon Tracker data that for the first
time have been correlated with TGFs detected by the AGILE Mini-Calorimeter.
We detect 8 TGFs with
gamma-ray photons of energies above 20 MeV  localized by the AGILE gamma-ray imager {\add  with an accuracy of $\sim
5-10^\circ$ at 50~MeV}.
Remarkably, all TGF-associated
gamma rays are compatible with a terrestrial production
site closer to the sub-satellite point than 400~km. 
Considering that our gamma rays reach the AGILE satellite at
 540 km altitude with  limited scattering or attenuation, our
measurements provide the first precise direct localization of TGFs
from space.
\end{abstract}

\pacs{}

\maketitle



\section{Introduction}

 Earth's atmospheric events associated with strong
thunderstorms have been in  recent years observed to be the site
of very efficient particle acceleration and gamma-ray emission at
MeV energies and above \cite{Fishman1994,Smith2005,Tsuchiya2007,Tsuchiya2009}. Of particular interest are the
so-called Terrestrial Gamma-Ray Flashes (TGFs)  that are very  short 
(lasting {\add up to }a few milliseconds) bursts of high-energy  photons above
100 keV, first  detected by the BATSE instrument on board the
Compton Observatory \cite{Fishman1994}.   TGFs have been
associated with strong thunderstorms mostly concentrated in the
Earth's equatorial and tropical regions \cite{Smith2005,Grefenstette2009}. 
{\add TGFs are widely believed to be produced by Bremsstrahlung in the atmospheric layers by a population of runaway electrons accelerated to relativistic energies by strong electric fields inside or above thunderclouds. The secondaries generated during the acceleration process can be accelerated as well driving an avalanche multiplication \cite{Gurevich1992}, commonly referred to as Relativistic Runaway Electron Avalanche (RREA). 
However, the RREA mechanism alone is not sufficient to explain the rich phenomenology of TGFs, especially the observed fluence, and there is no consensus yet on the underlying physical conditions, production sites, radiation efficiencies and maximal energies. An interesting possibility to overcome some of these difficulties is the relativistic feedback mechanism \cite{Dwyer2007,Dwyer2008}, which 
predicts an avalanche multiplication factor and characteristic discharge time compatible with the observed fluence and time profile of TGFs.}
{\add Concerning the maximal energy, } the original BATSE detection of TGFs up to a few
MeV \cite{Fishman1994,Nemiroff1997} was superseded by
the RHESSI detection up to 20 MeV \cite{Smith2005,Dwyer2005}. Recently, the AGILE satellite
showed that TGF spectrum extends well above 20 MeV \cite{Marisaldi2010} (Tavani et al., 2010, submitted to Nature),  
as confirmed also by the $Fermi$-GBM detector \cite{Briggs2010}.

AGILE \cite{Tavani2008b} is a mission of the Italian Space Agency
(ASI) dedicated to astrophysics in the gamma-ray energy range
30~MeV -- 30~GeV, with a monitor in the X-ray band 18~keV --
60~keV \cite{Feroci2007}, operating since April 2007 in a low inclination
($2.5^\circ$) Low-Earth Orbit at 540~km altitude. The AGILE Gamma-Ray Imaging Detector (GRID) is a pair-tracking telescope
based on a tungsten-silicon tracker \cite{Prest2003}. The imaging
principle is based on the reconstruction of the tracks left in the silicon
detection planes by the electron-positron pairs produced by
the primary photon  converting mainly in the tracker  tungsten planes. 
A Mini-Calorimeter
(MCAL) \cite{Labanti2009}, based on CsI(Tl) scintillating bars for
the detection of gamma-rays in the range 300~keV -- 100~MeV,
{\add and a plastic anti-coincidence detector \cite{Perotti2006}}
complete the high-energy instrument. MCAL can work also as an
independent gamma-ray transient detector with a dedicated
 trigger logic acting on several time scales
spanning four orders of magnitude between 290~$\mu$s and 8~seconds
\cite{Fuschino2008,Argan2004}. Thanks to its flexible trigger logic on {\add sub-millisecond} time scales, MCAL proved to be a very efficient instrument for TGF detection. The average MCAL detection rate is
$\sim 10$~TGFs/month, with the current selection criteria
\cite{Marisaldi2010}.

Up to now TGF observations have only been reported by space
instruments with no or quite limited on-board
imaging capabilities (e.g. BATSE and $Fermi$-GBM). 
The aim of this \textit{Letter} is
to provide a
first accurate {\add localization}  of TGFs from a space instrument and to study their very significant high-energy tail of
emission above 20~MeV. 
%

\section{AGILE-GRID  detection of TGFs}
\label{obs}

\begin{figure}
\includegraphics{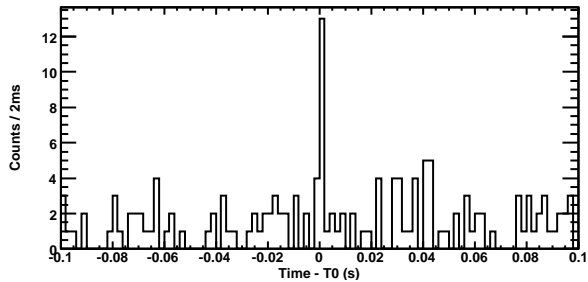}
\caption{\label{dt} Histogram of the arrival times with
respect  to the TGF MCAL-start time  $T_0$  for the gamma-rays detected
by the AGILE gamma-ray imaging detector (GRID).}
\end{figure}

In the period between June~2008 and December~2009 the MCAL
instrument triggered 119 bursts identified as TGFs according to
the selection criteria discussed in \cite{Marisaldi2010}. For each
of these bursts, the GRID  dataset was searched for
quasi-simultaneous gamma-ray events within a 200~ms time-window
centered at the TGF start time $T_0$, defined as the time of the first  MCAL-photon associated with the TGF. 
Figure \ref{dt} shows the  cumulative histogram of the arrival times of the
GRID events with respect to $T_0$  obtained by summing all the 119 bursts. A peak in the distribution is
evident for the 2~ms time bin immediately following $T_0$. This
peak includes 13 events, and the probability for it to be a
statistical fluctuation {\add (13 events or higher) is $6.5^. 10^{-10}$ }  if we assume
that GRID events are not correlated to TGFs and are distributed
according to the Poisson law with the measured average rate of
5.1~counts/s. 
 All {\add these} GRID events take place during the TGF emission time interval estimated from MCAL data only.

\begingroup
\squeezetable
\begin{table*}
\caption{\label{tab1}Selected events properties.}
\begin{ruledtabular}
\begin{tabular}{cccdddddddc}
\multicolumn{4}{c}{TGF informations}                                         & \multicolumn{2}{c}{Satellite position}  & \multicolumn{5}{c}{GRID event}                       \\
  Trigger    & Date & $T_0$\footnotemark[1]   &   \Delta T\footnotemark[2]   & \text{Lon.}  & \text{Lat.}     &  T_G\footnotemark[3] & \text{Type}\footnotemark[4]  & \text{Energy}  &  \phi\footnotemark[5]  &  D\footnotemark[6]           \\
  no.        &      & (UT)                    &  \text{(ms)}                 &  (^\circ)    &   (^\circ)      & \text{(ms)}              &                              & \text{(MeV)}   & (^\circ)               &  \text{(km)}       \\
\colrule
7467-3                    & 2008-10-03   &   09:54:56.460193  & 1.6 &       120.79      &      0.53  &  1.148 &  \text{R}  &  20_{-10}^{+20}  &  29  &  300\\
9669-1\footnotemark[7]    & 2009-03-08   &   09:41:27.957428  & 2.6 &       110.96      &     -2.33  &  1.966 &  \text{F}  &  80_{-40}^{+80}  &  28  &  280\\
9769-6\footnotemark[7]    & 2009-03-15   &   13:01:03.083206  & 1.0 &        28.88      &     -2.43  &  0.136 &  \text{F}  &  50_{-25}^{+50}  &  35  &  390\\
10042-11                  & 2009-04-03   &   20:39:15.460827  & 1.0 &       -75.86      &      2.46  &  0.395 &  \text{R}  &  60_{-30}^{+60}  &  14  &  130\\
11508-7\footnotemark[7]   & 2009-07-16   &   17:23:09.087689  & 2.0 &        12.06      &     -2.14  &  0.958 &  \text{F}  & 100_{-50}^{+100}  &  29  &  300 \\
12805-1                   & 2009-10-16   &   14:22:33.341584  & 1.0 &        24.23      &     -1.19  &  0.645 &  \text{R}  & 110_{-55}^{+110}  &  34  &  370\\
12809-19                  & 2009-10-16   &   20:44:55.148691  & 1.6 &       -66.23      &     -0.90  &  0.907 &  \text{R}  &  40_{-20}^{+40}  &  25  &  260 \\
12809-19                  & 2009-10-16   &   20:44:55.148691  & 1.6 &       -66.23      &     -0.90  &  1.171 &  \text{R}  &  50_{-25}^{+50}  &  24  &  240\\
12818-21                  & 2009-10-17   &   12:27:56.797752  & 0.8 &        22.05      &     -1.94  &  0.447 &  \text{R}  &  40_{-20}^{+40}  &  10  &  100\\
\end{tabular}
\end{ruledtabular}
\footnotetext[1]{Start time of the TGF, defined as the time of arrival of the first photon}
\footnotetext[2]{TGF duration}
\footnotetext[3]{Time difference between GRID event time and $T_0$}
\footnotetext[4]{F: the event direction intersects the Earth. R: the reversed event direction intersects the Earth.}
\footnotetext[5]{Angle between event direction and the Nadir. A $5.8^\circ$ uncertainty can be considered for an average 60~MeV photon energy.}
\footnotetext[6]{Distance of the event direction projected to the Earth surface from the satellite footprint. The uncertainty due to the error on direction reconstruction is between 50--90~km {\add for an average 60~MeV photon}.}
\footnotetext[7]{Albedo filtering disabled}
\end{table*}
\endgroup

A standard direction reconstruction of every GRID event is
performed on board by means of a GRID-adapted Kalman filter {\add \cite{Giuliani2006,Kalman1960}}: the
event is discarded if the incoming direction is found to be within
$70^\circ$ from the Earth center. This procedure,
called \textit{albedo filtering}, is aimed at the rejection of the
Earth gamma-ray albedo photons which are one of the most
significant contribution to the gamma-ray background.  This
procedure applies to all GRID events considered in our analysis,
except for the events obtained during a period of about 100 days
(1495 orbits) during which the albedo filtering was disabled for
test purposes.
%
For any given track in the GRID, {\add the \textit{albedo filtering}} assumes an incoming direction compatible
with the satellite field of view and cannot easily
discriminate whether  a photon producing a single or
quasi-single {\add (i.e. an event configuration of ambiguous topology)} track in the GRID originated  from the opposite
direction. 
This process is especially important  for low energy (few tens of MeV) photons which
tend to  produce single or quasi-single tracks with a few
detection planes involved. 

{\add 
All the selected GRID events were processed by the AGILE standard analysis pipeline which iteratively fits the tracks by means of a custom Kalman filter \cite{Kalman1960} in order to extract the most probable energy and direction of the incoming photon.
The nominal uncertainty on energy estimates is a factor $\sim 2$, mainly due to the fact that MCAL is only 1.5 radiation lengths thick and cannot provide full energy containment for electromagnetic showers. The angular resolution (68\% containment radius) is
$3.5^\circ$ at 100~MeV \cite{Tavani2008b} and scales approximately as the inverse of energy, being dominated by the multiple scattering effects which affect mostly the less energetic tracks. }
Using the
satellite position provided by the GPS system, it is checked
whether the event's incoming direction is compatible with  an
origin on the Earth surface, parameterized as the {\add World Geodetic System} WGS84
ellipsoid. If this is true, the event is flagged as a
\textit{Forward} (F) event. If no intersection is found, the event
is processed again with a non-standard algorithm that assumes
that the track vertex is
in the bottom planes of the tracker and the tracks
are allowed to develop from the bottom to the front planes. This
algorithm returns new estimates of energy and incoming direction.
It is then checked again whether the new incoming direction is
compatible with the Earth's surface. If this is the case,
the event is flagged as a \textit{Reverse} (R) event.
{\add Nine events out of 13 are classified as either F or R events. }
{\add For these 9 events, Table \ref{tab1} reports the
main properties of the TGFs and associated GRID events. It is worth noting that all the 3 $F$ events have been detected during observations with the \textit{albedo filtering} disabled.}
 The effects of a TGF production altitude {\add between 15--40~km \cite{Dwyer2005,Ostgaard2008}} have not been considered at this stage because they are much smaller than the uncertainties due to the instrument angular resolution.
{\add Among the four GRID events with directions not compatible
with the Earth's surface, three of them come from directions very close to the Earth's limb, which make them classifiable as
albedo photons passing the on-board \textit{albedo filtering} step.} 

\begin{figure}
\includegraphics[width=7.5cm]{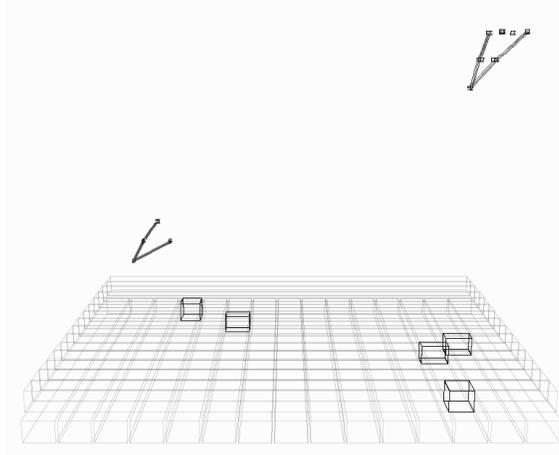}
\caption{\label{track3D} 3D view of the tracks
reconstructed for the two GRID events detected in association with TGF~12809-19.  The MCAL detector with the detected events is shown at the bottom of the figure. The AGILE pointing direction is orthogonal to the MCAL detection planes, toward the top of the figure. 
}
\end{figure}

{\add All TGFs in the selected sample have one associated GRID event exept} TGF~12809-19, for which
two closely spaced GRID events were detected. For this TGF, Fig.
\ref{track3D} shows the topology of the tracks associated to the
two photons detected in the tracker.
{\add 
The tracks shown join the hits detected in the silicon tracker planes (the planes are not shown for clarity). For both events, a track vertex can be found indicating that the events are coming from the bottom of the instrument. 
}

\section{Results and discussion}

The  incoming directions of the 9 selected events appear to be clustered close
to the subsatellite point, with an average (maximum) $\phi$ angle {\add (the angle between the photon direction and the satellite nadir)}
of $25.4^\circ$ ($35.1^\circ$) {\add and} distance {\add to the subsatellite point} of {\add 260~km (390~km)}. 
All 9 events are thus contained within a 1.14~sr solid angle, a
factor 3.4  smaller than the solid angle subtended by the Earth at
the satellite altitude of 540~km,  which corresponds to a maximum visibility projected distance radius of $\sim 2600$~km from the satellite footprint. 

\begin{figure*}[t]
\includegraphics{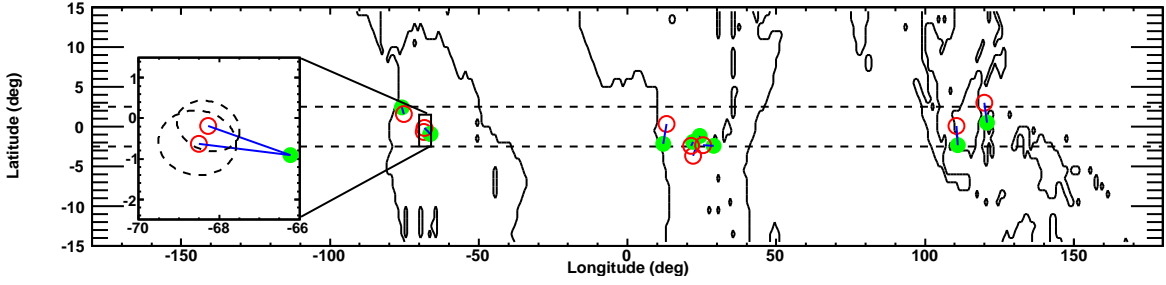} 
\caption{\label{map} Location of the satellite footprints (filled circles) and of the projection of the associated tracker events to the Earth ellipsoid (hollow circles). The inset is a zoom over the region of TGF~12809-19. The dashed ellipses represent the $1\sigma$ error circle for the associated tracker events.}
\end{figure*}

\begin{figure}
\includegraphics[width=8.5cm]{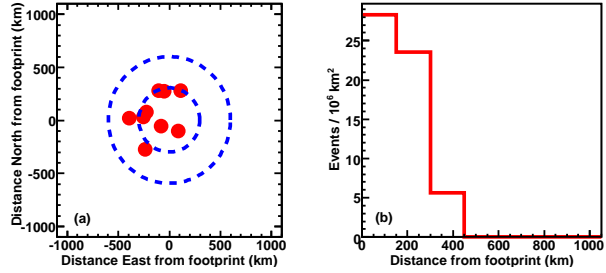}
\caption{\label{dist} (a) scatterplot of the GRID events projection with respect to the AGILE footprint. The dashed circles are 300 and 600~km in radius. The full AGILE visibility region has a 2600~km radius. (b) occurrence density vs. distance from footprint (each bin has been divided by the subtended area in $\mathrm{km}^2$).}
\end{figure}

Figure \ref{map} shows the
geographical location of the satellite for the considered TGFs as
well as the location of the intersection of the tracked events
direction with the Earth. It is remarkable to note that the two GRID events associated to 
TGF~12809-19, both with energies of about 40~MeV, come
from very close directions, separated only by a $5.4^\circ$ angle,
within the instrument angular resolution at these energies. The
time delay between the two events is only 264~$\mu$s. Considering
that the dead-time for GRID data acquisition is 200~$\mu$s, it is
possible that high energy measurements for this TGF, as well as
for others, is hampered by the dead-time.
 Figure \ref{dist} shows the scatterplot of the GRID events projection with respect to the AGILE footprint, and the distribution of the occurrence density vs. distance from footprint (each bin has been divided {\add by} the subtended area in $\mathrm{km}^2$). 

 Gamma-ray photons can be Compton scattered in their path from
the TGF source to the satellite. We calculated the scattering
probability and angular distribution of the resulting gamma-ray
flux for a TGF source at 
{\add 15-40~km altitude \cite{Dwyer2005,Ostgaard2008}, using the photon cross-sections provided by the NIST XCOM database\cite{Xcom2009} and assuming an air density vertical profile scaling exponentially with a length scale of 7~km. Above 40~km, the probability for a photon of energy larger than 20~MeV to be Compton scattered results lower than $3\%$. If a Compton interaction takes place in the 15-40~km altitude interval, the resulting positional smearing will be
within the angular resolution of the GRID reconstruction, considering also that small scattering angles are favoured at high energy. }
Furthermore, we also considered the possibility of the production
of secondaries from below originating either {\add by Compton scattering or pair production}
in the spacecraft or in MCAL, {\add which can contribute to our R-detections}. 
In both cases, the resulting angular smearing is of the same order of the GRID angular resolution.

 Our detection of TGFs with an imaging gamma-ray detector is
important for several reasons: (1) it provides the first accurate
{\add localization} (within a few degrees) of TGFs from space; (2) it
reinforces the AGILE discovery that a significant high-energy
component of emission is produced by TGFs well above 20~MeV; (3)
it shows that the TGF high-energy emission is detected by
satellites in LEO orbits from a relatively small region within
300-400 km from the satellite footprint. 
{\add The AGILE detection of gamma-ray photons above 50~MeV from TGFs (the average energy of the 9 GRID events is 60~MeV) strongly constrains the theoretical models. For example, the detection of a given energy sets a lower limit in the electric potential difference involved. In the relativistic feedback scenario \cite{Dwyer2007}, where the maximum possible static electric field is limited by the avalanche mechanism, this correspond to a lower limit in the column depth of the avalanche region hence, given a production altitude, sets a lower limit in the extension of the avalanche region. 
Moreover,  high energy electrons are expected to be very well aligned with the electric field \cite{RousselDupre2008}. Considering the small angular scattering of Bremsstrahlung photons for highly relativistic electrons, the incoming direction of high energy photons marks well the electric field orientation at the source. In this sense our localization technique provides also a diagnostic tool for the electric field at the source region. }
{\add The observed clustering of TGFs close to the subsatellite point} is in agreement with other independent
determinations of TGF locations based on ground measurements {\add (sferics)} simultaneously obtained with space
detections {\add by the RHESSI satellite}  \cite{Cummer2005,Hazelton2009,Cohen2010}. 
Future investigations will
determine  whether the apparently narrow cone of  detected gamma-rays
 is  due to beamed emission intrinsic to the TGF source or caused by a selection
effect favored by absorption and Compton scattering in the
atmosphere.


\begin{acknowledgments}
AGILE is a mission of the Italian Space Agency (ASI), with
co-participation of INAF (Istituto Nazionale di Astrofisica) and
INFN (Istituto Nazionale di Fisica Nucleare).  Research
partially funded through the ASI contract n. I/089/06/2.
\end{acknowledgments}


%

\end{document}